\documentclass[sigconf]{acmart}
\usepackage{booktabs} 

\copyrightyear{2018}
\acmYear{2018}
\setcopyright{acmlicensed}
\acmConference[AVI '18]{2018 International Conference on Advanced Visual Interfaces, AVI '18}{May 29-June 1, 2018}{Castiglione della Pescaia, Italy}
\acmBooktitle{AVI '18: 2018 International Conference on Advanced Visual Interfaces, AVI '18, May 29-June 1, 2018, Castiglione della Pescaia, Italy}
\acmPrice{15.00}
\acmDOI{10.1145/3206505.3206516}
\acmISBN{978-1-4503-5616-9/18/05}

\begin{document}
\title{Map-Based Visualization of 2D/3D Spatial Data via Stylization and Tuning of Information Emphasis
}

\author{Liliana Ardissono, Matteo Delsanto, Maurizio Lucenteforte, Noemi Mauro, Adriano Savoca and Daniele Scanu}
\affiliation{%
  \institution{Computer Science Department, University of Torino}
  \streetaddress{Corso Svizzera 185}
  \city{Torino} 
  \state{Italy} 
  \postcode{I-10149}
}
\email{liliana.ardissono,maurizio.lucenteforte,noemi.mauro@unito.it}






\renewcommand{\shortauthors}{Ardissono et al.}
\renewcommand{\shorttitle}{Stylization and Tuning of Information Emphasis}

\begin{abstract} 
In Geographical Information search, map visualization can challenge the user because results can consist of a large set of heterogeneous items, increasing visual complexity.
We propose a novel visualization model to address this issue. Our model represents results as markers, or as geometric objects, on 2D/3D layers, using stylized and highly colored shapes to enhance their visibility. Moreover, the model supports interactive information filtering in the map by enabling the user to focus on different data categories, using transparency sliders to tune the opacity, and thus the emphasis, of the corresponding data items. 
A test with users provided positive results concerning the efficacy of the model.
\end{abstract}

\ccsdesc[200]{Information systems~Geographic information systems}
\ccsdesc[200]{Human-centered computing~Visualization}
\ccsdesc[200]{Information systems~Search interfaces}

\keywords{Search Results Visualization; 2D/3D Geographical Maps; Opacity Tuning; Visual Information Filtering}

\maketitle

\section{Introduction}

Geographical maps can challenge information search in various ways. For instance, the queries of a search session can focus on multiple data categories and search results should be simultaneously visualized in a map to provide a unified access point to information. However, a large number of items might clutter the map, increasing its visual complexity. 
Moreover, geographical results have a geometry that cannot be neglected (e.g., rivers and parks), or that deserves representation in order to make them recognizable in the environment; e.g., consider specific buildings within 3D city maps. Therefore, the visualization cannot be limited to the presentation of markers: geometries should be displayed as well. 
 
In this paper we propose a novel visualization model aimed at addressing the above issues. Our model supports the management of 2D and 3D maps and is characterized as follows:
\begin{itemize}
\item
{\em Stylization and highlighting of search results.}
The model employs vivid colors and stylized shapes to graphically distinguish the searched items from the other elements included in the background layers of the maps, and to represent data types; i.e., the categories to which information items belong; e.g., schools vs. hospitals. Different from Virtual Globes, which mirror reality in a realistic way, we exploit graphical difference to make data evident to the user. 
\item
{\em Interactive information filtering through opacity tuning.}
In order to enable the user to emphasize/de-emphasize information in a selective way, we introduce a typed interactive filtering model to modify data visualization without changing the content of the map.
Specifically, we propose transparency sliders for tuning the opacity of the visualized information, by data category. In that way, the user can steer visualization directly on the map, and he or she can decide which data categories have to be de-emphasized, and how much, in order to let the most relevant data emerge without loosing reference to the other search results. 
\end{itemize}
These features are aimed at graphically highlighting differences in types of information (using custom icons for different types) and at supporting abstraction, through information filtering; see \cite{Card-etal:99}.

We carried out a laboratory test with users to evaluate our visualization model in the OnToMap Participatory GIS (PGIS) \cite{Ardissono-etal:17b,Ardissono-etal:17d,Voghera-etal:16}. Our main research questions are the following:
\newline
{\em {\bf RQ1:} Does the emphasis of geographical information, on a 2D/3D map, help the user identify the data he or she needs during a Geographical Information search task?}
\newline
{\em {\bf RQ2:} Can opacity tuning help the user focus on relevant information during the analysis of geographical search results, with respect to visualizing data with a fixed opacity, the same for every category?}

The results of our experiments show that emphasizing search results helps people find the information they need. Moreover, opacity tuning is a powerful function to focus map content in order to satisfy detailed information needs; e.g., to temporarily focus on a subset of the search results relevant to the completion of a sub-task.

\section{Background and Related Work}
\label{background}

In \cite{Hu-etal:15}, Hu et al. point out that "empirical studies show that visualization technologies, such as 2D maps and 3D virtual environments, can facilitate participants' learning and understanding in decision-making, especially spatial decision-making, processes"; e.g., see \cite{Al-Kodmany:99,Simpson:01}. 
2D and 3D visualization were used in Web Collaborative GIS; e.g., see \cite{Hunter-etal:12}), and \cite{Sun-li:16,Hu-etal:15,Isikdag-Zlatanova:00}. In general, it is not clear whether 3D maps are superior to 2D ones in different application domains. \cite{Laakso-etal:03} reported that, in a tourist and navigation support service for mobile devices, 3D maps had advantages over 2D ones, but they might not provide much benefit for experienced 2D map users.

Concerning map readability and learnability,
Canham and Hegarty \cite{Canham-Hegarty:09} advocate minimality in graphical interface design: "graphics should not display more information than is required for the task at hand". 
Earlier, \cite{Allen-etal:06} built on Hegarty's prior work and reported that "individual differences in the ability to learn from simple maps, figures, and diagrams are a product of both domain-specific knowledge and general visual spatial abilities". Moreover, the addition of perceptual detail in a navigation interface (route map) impacts on map learning, depending on the user's spatial abilities \cite{Sanchez-Branaghan:09}.  
Recent work investigates the adaptation of information visualization to the user's characteristics \cite{Lalle-etal:17}. 

Some works attempt to reduce the complexity of geographical maps through abstraction. E.g., \cite{Wang-etal:14} proposes hierarchical route maps representing less or more detailed views. \cite{Dunlop-etal:07} varies the width of linear geometries (runs of pistes) to highlight the most relevant results. Other works exploit transparency to overlay different types of information on maps \cite{Luz-Masoodian:14}, to combine an attribute setting mechanism with the visualization of a background working area \cite{Harrison-etal:95}, to merge maps in an overlay model \cite{Elias-etal:08}, or to provide translucent layers for map exploration \cite{Lieberman:94}. In comparison, we employ transparency to enable the user to focus maps on subsets of information.

Visual interfaces are adopted in information retrieval to provide overviews \cite{Hornbaek-Hertzum:11} and help the comprehension of information; e.g., \cite{Andrienko-Andrienko:09,Angelini-etal:13}. Some works attempt to improve the presentation of results by displaying them in 3D maps, explicitly representing geographical, temporal, semantic, or other types of relations; e.g., \cite{Deeswe-Kosala:15,Sen-etal:17,Kunkel-etal:17}.  
Other works reduce visual complexity through sketching \cite{Wood-etal:12}, which seems to support user engagement.
We employ 2D/3D maps to represent the geographic extension of information, using a symbolic, stylized representation of data categories, in the tradition of Parish Maps and community mapping \cite{Parker:06}. 
However, the shapes of our model do not recall handy sketching.

\section{Visualization Model}
\label{model}

Our model assumes that data is semantically modeled in categories representing types of geographic information. The domain conceptualization is based on an OWL ontology representing different dimensions of the territory, among which the artificial and the natural one, and going to the level of detail of specific data types, such as schools, hospitals, etc.; see \cite{Voghera-etal:16,Mauro-Ardissono:17b,Ardissono-etal:16}.

\begin{figure}
 \includegraphics[width=1\columnwidth]{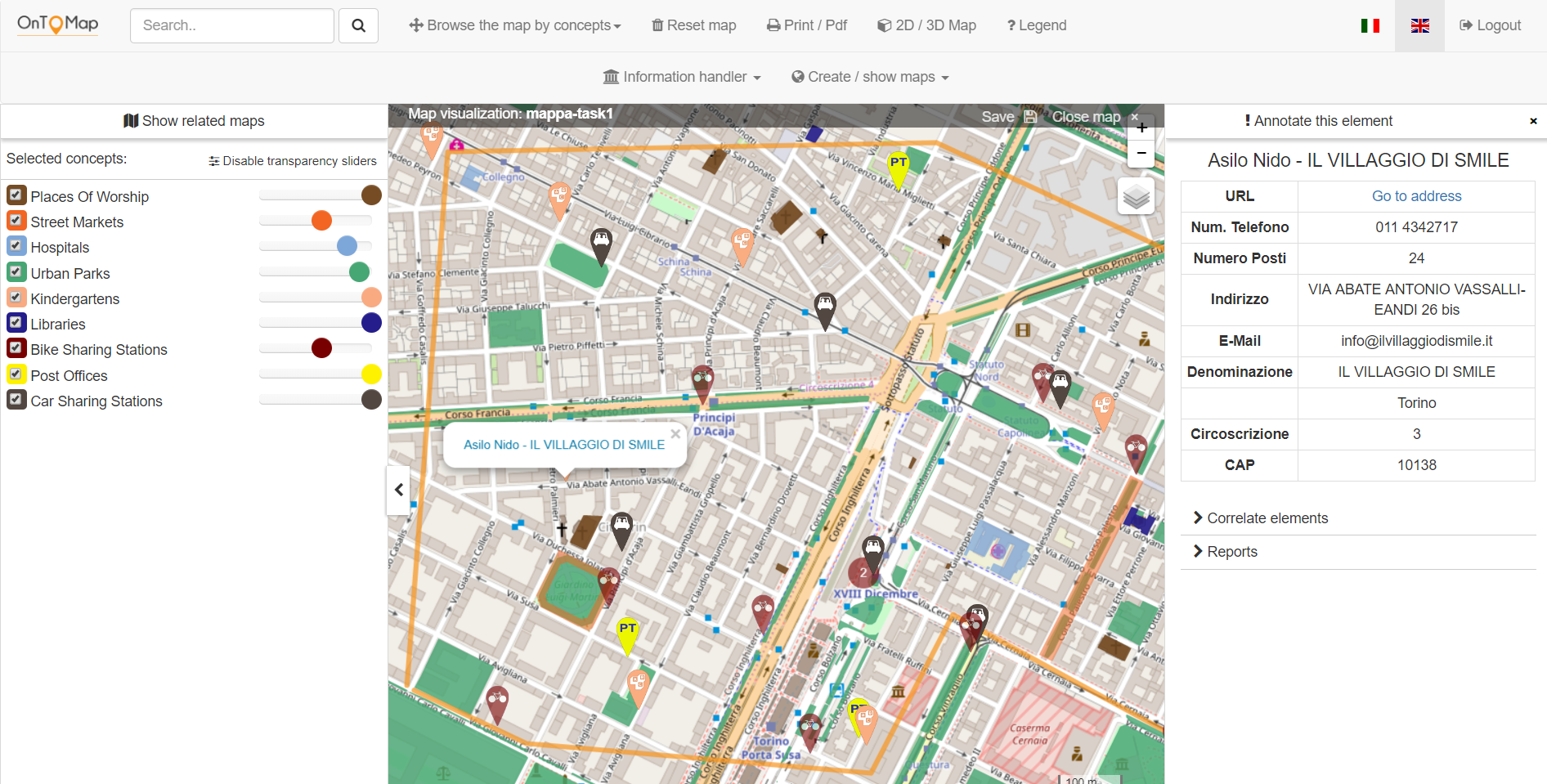}
 \caption{2D map generated by OnToMap.}
 \label{f:mappa2D}
 \end{figure}

Figure \ref{f:mappa2D} shows our visualization model for 2D maps, applied to the OnToMap PGIS \cite{Ardissono-etal:17d,Ardissono-etal:17b}.
The top of the page includes a textual search bar and the main commands for managing the maps.
The map shows search results in the current bounding box. Geo-data is depicted either as markers (if the geometry is not available) or as shapes (in the opposite case) in vivid colors, in order to enhance item identification; the colors of items depend on their data categories. 
The tab in the left portion of the page shows the data categories selected by the user as checkboxes with transparency sliders; as above, colors depend on data categories. 
The transparency sliders support the tuning of the opacity of the corresponding data items. For instance, in the figure, the street markets, hospitals, urban parks and bike sharing stations are semi-transparent.

By clicking on a marker, or on the shape of an item, the user can view a table providing details about it; see Figure \ref{f:mappa2D}. 

\begin{figure}
 \includegraphics[width=1\columnwidth]{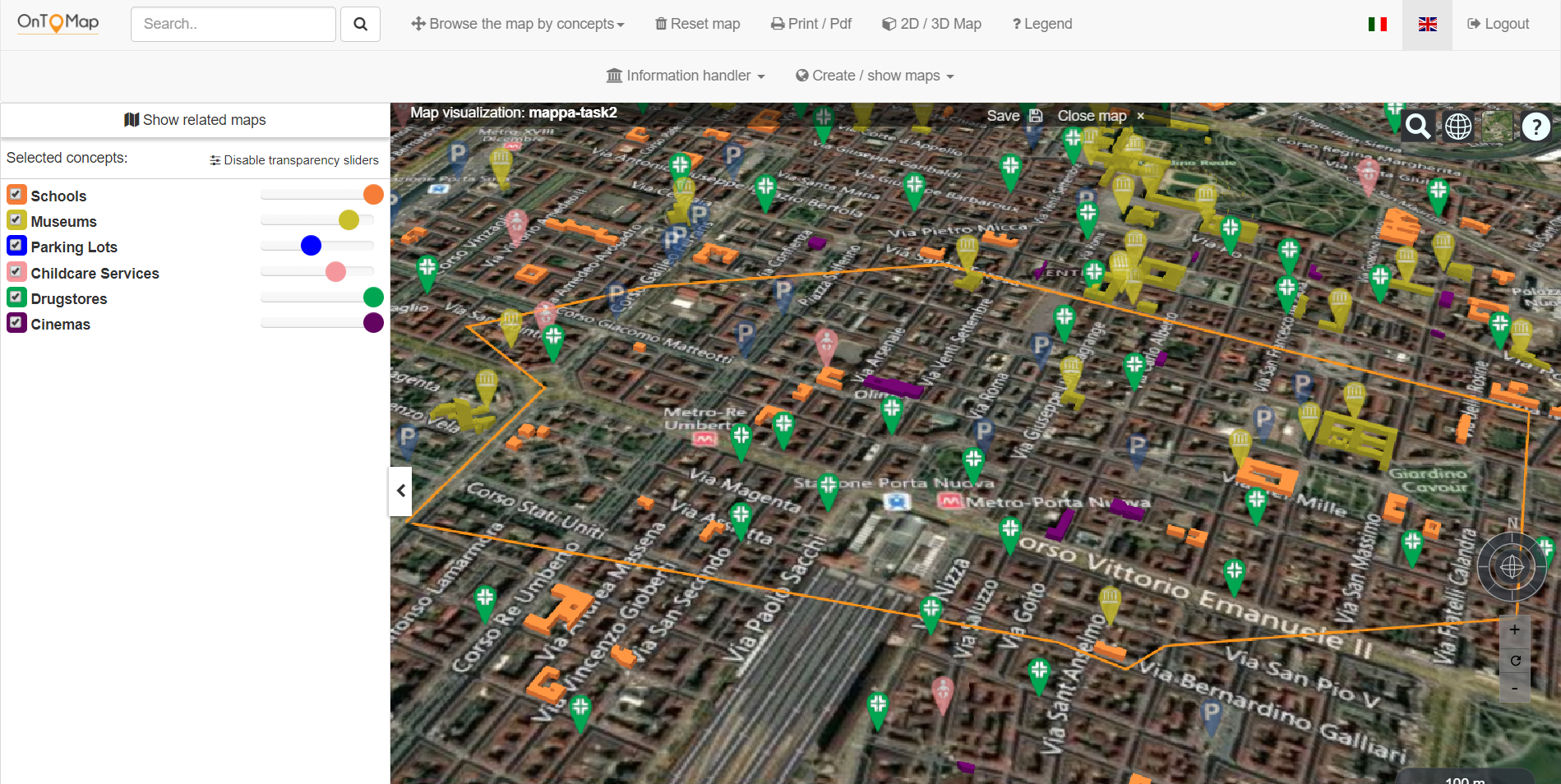}
 \caption{3D map generated by OnToMap.}
 \label{f:mappa3D}
  \end{figure}

Figure \ref{f:mappa3D} shows the 3D map, which has the same features as the 2D one. Search results are depicted as solid, stylized, vividly colored shapes, and are overlaid on the 3D terrain layer: they cover the corresponding objects, but they are stylistically different for discernibility purposes.

\section{Technical Details}
\label{technical}

The user interface of OnToMap is implemented using HTML5 mark-up language + CSS, and uses two different libraries for embedding dynamic interactive maps: Leaflet (http://leafletjs.com/), that supports 2D visualization based on OpenStreetMap, and Cesium (https://cesiumjs.org/), a JavaScript library for 3D globes offering different maps as a base layer (e.g., Bing, OpenStreetMap, etc.). 

Given a search query, the back-end of the application retrieves the results from the semantic layer as a GeoJSON (http://geojson.org/) formatted dataset (a {\em FeatureCollection}), representing a structured set of GeoJSON objects, each one containing geographic coordinates and geometry data of an item belonging to that category. These objects are visualized on the 2D or 3D map elements as polygons (or multi-polygons), lines (or multi-lines) or points, represented with colors and icons depending on their own category. 

As the obtained dataset represents a set of two-dimensional geometries, it is necessary to manipulate each of its entities before passing it to the 3D viewer. In particular, a type check (polygon, polyline or billboard) is carried out. Depending on the specific type, the properties values are used to assign the appropriate color to the {\em material} field, a value representing the height in meters to the {\em extruded-Height} field (only for polygons) and a  link to the image in PNG format to be displayed in the marker indicating the point geometry to the image field (only for billboards). At this point, entities can be added to the scene and displayed by the current viewer: focusing on 3D maps (2D visualisation is handled in a similar way using Leaflet), the dataset items will be added to the Cesium viewer and displayed on the map.

The opacity of items, tuned by moving the sliders, is rendered in 2D by  modifying the properties of the CSS. In 3D maps, the color {\em alpha} channel of the item object is updated using the Model-View-ViewModel (MVVM) pattern.

\section{Visualization Model Validation}
\label{experiment}

We evaluated our model in a laboratory test with users, using OnToMap (OTM), in 2D and 3D modalities, and comparing it with Google Maps (GM) 2D and 3D, representing the baselines.
The experiment had a single-factor, within-subjects design. The investigated factor was the role of emphasis (expressed as stylization, vivid colors and through opacity) in map-learning, in 2D and 3D settings.

We asked participants to perform 4 map-learning tasks, each one associated with a different map: 
\begin{itemize}
\item 
Task1: question answering on a 2D map generated by OTM;
\item
Task2: the same, but on a 3D map generated by OTM;
\item
Task3: the same on a 2D map generated by GM;
\item  
Task4: the same on a 3D map generated by GM. 
\end{itemize}
In order to compare 2D or 3D maps across applications, we considered each treatment condition as an independent variable and every participant received the 4 treatments. We counterbalanced the order of tasks to minimize the effect of practice and fatigue. 

\subsection{Test Subjects}
For the experiment we recruited 54 people (48,1\% females and 51,9\% males) who worked for free, without any reward. They are part of the university staff (teachers and secretaries) and students, as well as people working in the industry or retired.
They have different backgrounds (52.9\% scientific, 15.7\% technical, 19.6\% humanities and linguistics, 5.8\% arts, 5.8\% social sciences and law). Their age is between 20 and 75 years, with a mean age = 32.04.
Most of the participants use geographical maps sometimes, or often, at work, at home, or during their free time. However, they are more used to 2D maps than 3D ones: the 75.9\% of participants declared that they often use 2D maps but only 5.6\% declared the same of 3D maps. Most people rarely (57.4\%) or never (37\%) use 3D maps.

\subsection{The experiment}

One person at a time performed the test, which lasted about 40 minutes. 
Participants first used OnToMap and Google Maps for a few minutes, in 2D and in 3D modalities, in order to get acquainted with the two types of maps. 
Then, we instructed them about the tasks to be performed. For each task, we provided them with one of the four maps and with the questions to be answered looking at the visualized information.

In order to focus on map-learning, the maps were populated with the data to be analyzed. Information consisted of instances of multiple data categories.
Each map was annotated with an orange line that delimited the geographical area to be analyzed.
Figures \ref{f:mappa2D} and \ref{f:mappa3D} show the OTM maps we used; we omit the GM ones for brevity. The maps differed from one another in bounding box and in the visualized data categories, to prevent learning effects during task execution. 

For each task, we asked participants to answer 4 questions (2 for each template below) using the interactive functions of the map: 
\begin{enumerate}
\item
How many \textit{category name} (e.g., hospitals) are visualized within the area delimited by the orange line of the map?
\item
Find item {\em Y} (e.g., "Il villaggio di Smile") within the orange line in the map, and visualize detailed information about it.
\end{enumerate}
As objective performance indicators, we measured task completion time and the percentage of correctly answered questions.
As a subjective measure, we analyzed user experience: at the end of each task, participants answered a questionnaire aimed at evaluating the map they had just used; see Table \ref{t:questions}. They provided values in a 7-point Likert scale from 1, the worst value, to 7, the best one.

After the completion of all the tasks, participants filled in a post-test questionnaire eliciting feedback about the best and worst characteristics of OnToMap 2D and 3D. 

\begin{table}
\centering
\caption{Questions proposed after the execution of each of the 4 tasks (translated from the Italian language)}
\begin{tabular}{|l|l|l|}
\hline
\textbf{\#} & \textbf{Task} & \textbf{Question} \\ \hline
1 & 1/2/3/4 & Please, rate the ease of use of the map.   \\ \hline
2 & 1/2/3/4 & How easily did you find the information \\
  &         & you were looking for in the map? \\ \hline
3 & 1/2/3/4 & How clear was the visualization \\
  &         & of search results in the map?   \\ \hline
4 & 1/2/3/4 & Please, rate the attractiveness of the map.   \\ \hline
5 & 1/2/3/4 & Please, rate the novelty of the map.   \\ \hline
6 & 1/2 & Please, evaluate the usefulness of the \\
  &     & transparency slider in information filtering. \\ \hline 
\end{tabular}
\label{t:questions}
\end{table}


\begin{table}
\centering
\caption{Participants' performance during task execution}
\begin{tabular}{ccccc}
\hline
\multicolumn{1}{|c|}{{\bf Task}} & 
\multicolumn{1}{c|}{\textbf{\begin{tabular}[c]{@{}c@{}}Min\\  Time\end{tabular}}} & 
\multicolumn{1}{c|}{\textbf{\begin{tabular}[c]{@{}c@{}}Max\\  Time\end{tabular}}} & 
\multicolumn{1}{c|}{\textbf{\begin{tabular}[c]{@{}c@{}}Median\\  Time\end{tabular}}} & 
\multicolumn{1}{c|}{\textbf{\begin{tabular}[c]{@{}c@{}}Mean \% of \\ correct answers\end{tabular}}}  \\ \hline
\multicolumn{1}{|c|}{\textbf{1: OTM 2D}} & \multicolumn{1}{c|}{2}                                                                          & \multicolumn{1}{c|}{{\bf 238}} 
& \multicolumn{1}{c|}{18}                                                                        & \multicolumn{1}{c|}{87.96\%}                     \\ \hline
\multicolumn{1}{|c|}{\textbf{2: OTM 3D}} & \multicolumn{1}{c|}{3}                                                                          & \multicolumn{1}{c|}{170}                                                                         
& \multicolumn{1}{c|}{{\bf 22}}                                                                        & \multicolumn{1}{c|}{93.52\%}                     \\ \hline
\multicolumn{1}{|c|}{\textbf{3: GM 2D}} & \multicolumn{1}{c|}{2}                                                                          & \multicolumn{1}{c|}{85}                                                                         
& \multicolumn{1}{c|}{11}                                                                        & \multicolumn{1}{c|}{{\bf 98.61\%}}                     \\ \hline
\multicolumn{1}{|c|}{\textbf{4: GM 3D}} & \multicolumn{1}{c|}{3}                                                                          & \multicolumn{1}{c|}{71}                                                                              
& \multicolumn{1}{c|}{10}                                                                        & \multicolumn{1}{c|}{87.96\%}                     \\ \hline
\multicolumn{1}{l}{}                 & \multicolumn{1}{l}{}                                                                            & \multicolumn{1}{l}{}                                                                               
& \multicolumn{1}{l}{}                                                                           & \multicolumn{1}{l}{}                            
\end{tabular}
\label{t:time_answers}
\end{table}

\section{Evaluation Results}
\label{evaluation}

\subsection{User performance}
\label{userPerformance}

Table \ref{t:time_answers} shows the execution time of each task, in seconds. The last column shows the percentage of correct answers per task.

We think that Task1 and Task2 were longer than the other tasks because the maps generated by OTM provide more details about geographical items; thus, it took more time to read them. Moreover, participants spent time using the transparency sliders to dynamically include and exclude data categories during information search.  

Regarding the correct answers, Task3 achieved the best score, probably because (i) the familiarity of people with the icons used by Google Maps helped information finding; (ii) in the OTM 2D map, some data categories were difficult to discern because they were displayed in colors similar to the background layer.

\begin{table}[]
\centering
\caption{Statistics about Task1 and Task3}
\begin{tabular}{|l|c|c|c|c|c|c|}
\hline \hline
\multicolumn{7}{|c|}{\textit{\textbf{Task1 (OTM 2D)}}}                                                                                                                                                                                                                                                                                                                                                                                                                \\ \hline \hline
\textbf{Question \#} & \textbf{\begin{tabular}[c]{@{}c@{}} 1 \end{tabular}} & \textbf{\begin{tabular}[c]{@{}c@{}} 2 \end{tabular}} & \textbf{\begin{tabular}[c]{@{}c@{}} 3 \end{tabular}} & \textbf{\begin{tabular}[c]{@{}c@{}} 4\end{tabular}} & \textbf{\begin{tabular}[c]{@{}c@{}} 5\end{tabular}} & \textbf{\begin{tabular}[c]{@{}c@{}} 6\end{tabular}} \\ \hline
\textbf{Mean}     & \underline{\textbf{6.07}}      & \underline{\textbf{5.37}}                                                                                                 & 5.11                                                                                                            & \textbf{5.52}                                                                & \textbf{5.57}    & \underline{\textbf{5.96}}                                                                                 \\ \hline
\textbf{Variance} & 1.13 & 1.82                                                        & 2.67                                                                                                            & 1.88                                                                         & 1.00             & 3.17                                                                                 \\ \hline
\textbf{St. Dev.} & 1.06               & 1.35                                                                                                          & 1.63                                                                                                            & 1.37                                                                         & 7.00             & 1.78                                                                                 \\ \hline
\hline

\multicolumn{7}{|c|}{\textit{\textbf{Task3 (GM 2D)}}}                                                                                                                                                                                                                                                                                                                                                                                                            \\ \hline \hline
\textbf{Question \#} & \textbf{\begin{tabular}[c]{@{}c@{}} 1 \end{tabular}} & \textbf{\begin{tabular}[c]{@{}c@{}} 2 \end{tabular}} & \textbf{\begin{tabular}[c]{@{}c@{}} 3 \end{tabular}} & \textbf{\begin{tabular}[c]{@{}c@{}} 4\end{tabular}} & \textbf{\begin{tabular}[c]{@{}c@{}} 5\end{tabular}} & \textbf{\begin{tabular}[c]{@{}c@{}} 6\end{tabular}} \\ \hline
\textbf{Mean}     & 5.67               & 4.96                                                                                                          & \textbf{5.35}                                                                                                   & 5.19                                                                         & 3.72             & \multicolumn{1}{l|}{}                                                                \\ \hline
\textbf{Variance} & 1.36               & 2.41                                                                                                          & 2.27                                                                                                            & 2.53                                                                         & 2.54             & \multicolumn{1}{l|}{}                                                                \\ \hline
\textbf{St. Dev.} & 1.17               & 1.55                                                                                                          & 1.51                                                                                                            & 1.59                                                                         & 1.60             & \multicolumn{1}{l|}{}                                                                \\ \hline
\end{tabular}
\label{t:2D}
\end{table}

\begin{table}[]
\caption{Statistics about Task2 and Task4}
\centering
\begin{tabular}{|l|c|c|c|c|c|c|}
\hline \hline
\multicolumn{7}{|c|}{\textit{\textbf{Task2 (OTM 3D)}}}                                                                                                                                                                                                                                                                                                                                                                                                                \\ \hline \hline
\textbf{Question \#} & \textbf{\begin{tabular}[c]{@{}c@{}} 1 \end{tabular}} & \textbf{\begin{tabular}[c]{@{}c@{}} 2 \end{tabular}} & \textbf{\begin{tabular}[c]{@{}c@{}} 3 \end{tabular}} & \textbf{\begin{tabular}[c]{@{}c@{}} 4\end{tabular}} & \textbf{\begin{tabular}[c]{@{}c@{}} 5\end{tabular}} & \textbf{\begin{tabular}[c]{@{}c@{}} 6\end{tabular}} \\ \hline
\textbf{Mean}     & \textbf{5.83}      & \textbf{5.28}                                                                                                 & \underline{\textbf{5.57}}                                                                                                   & 5.09                                                                         & \underline{\textbf{5.76}}    & 5.67                                                                                 \\ \hline
\textbf{Variance} & 1.65               & 2.13                                                                                                          & 1.80                                                                                                            & 3.33                                                                         & 2.00             & 4.04                                                                                 \\ \hline
\textbf{St. Dev.} & 0.28               & 1.46                                                                                                          & 1.34                                                                                                            & 1.83                                                                         & 7.00             & 2.01                                                                                 \\ \hline 
\hline
\multicolumn{7}{|c|}{\textit{\textbf{Task4 (GM 3D)}}}                                                                                                                                                                                                                                                                                                                                                                                                            \\ \hline \hline
\textbf{Question \#} & \textbf{\begin{tabular}[c]{@{}c@{}} 1 \end{tabular}} & \textbf{\begin{tabular}[c]{@{}c@{}} 2 \end{tabular}} & \textbf{\begin{tabular}[c]{@{}c@{}} 3 \end{tabular}} & \textbf{\begin{tabular}[c]{@{}c@{}} 4\end{tabular}} & \textbf{\begin{tabular}[c]{@{}c@{}} 5\end{tabular}} & \textbf{\begin{tabular}[c]{@{}c@{}} 6\end{tabular}} \\ \hline
\textbf{Mean}     & 5.74               & 5.00                                                                                                          & 5.26                                                                                                            & \underline{\textbf{5.94}}                                                                & 4.43             & \multicolumn{1}{l|}{}                                                                \\ \hline
\textbf{Variance} & 1.78               & 1.92                                                                                                          & 2.38                                                                                                            & 1.22                                                                         & 2.44             & \multicolumn{1}{l|}{}                                                                \\ \hline
\textbf{St. Dev.} & 1.33               & 1.39                                                                                                          & 1.54                                                                                                            & 1.11                                                                         & 1.56             & \multicolumn{1}{l|}{}                                                                \\ \hline
\end{tabular}
\label{t:3D}
\end{table}

\subsection{User experience}

Table \ref{t:2D} shows the statistics concerning the answers to the questionnaires proposed after Task1 and Task3 (column titles denote the questions in Table \ref{t:questions}). 
In the comparison between OTM 2D and GM 2D, the OTM maps received the best evaluations in terms of ease of use ($\alpha=0.05$), ease of identification of information ($\alpha=0.07$), novelty ($\alpha=0.05$) and attractiveness ($\alpha>0.1$). GM got the highest rating for the clarity of information visualization, in line with the ratio of correct answers in Table \ref{t:time_answers}. We attribute the positive evaluation of the ease to find information in OTM to the transparency sliders, which helped item identification by simplifying the map. In contrast, the lower evaluation of visualization clarity can be related to the above mentioned color issue.

Table \ref{t:3D} shows the results concerning Tasks 2 and 4. 
OTM 3D outperformed GM 3D in all the measures but the attractiveness of the map, which was probably challenged by the lower definition of OTM background layer w.r.t. that of GM. However, the only significant results concern attractiveness and novelty ($\alpha=0.05$).

In both tables, the boldface, underlined numbers are the best values in the comparison among all of the four maps. We can see that OTM 2D has the highest ease of use and ease of identification of information;
OTM 3D has the highest clarity of visualization and novelty;
Google Maps 3D wins in attractiveness. 

Columns 6 of Tables \ref{t:2D} and \ref{t:3D} show that participants considered the transparency sliders as more useful in the 2D map than in the 3D one. This can be explained by the superior clarity of the 3D map, in which the vivid colors of geographical items were very evident on the background layer, and helped information finding.

The post-test questionnaire confirmed that the best feature of OTM was the information filtering support provided by transparency sliders. 
People appreciated the visualization of geometries, especially in 3D, because they help recognizing buildings in the city.
People also liked the representativity of the icons of markers because they help recognizing the type of information on the map.

The worst characteristics of OTM concerned:
(i) The colors used to represent some data categories, which in the 2D maps reduced visibility.  
(ii) The fact that information items having small geometries are hard to identify. Some users suggested to combine shapes with markers to enhance visibility and to help understand if buildings having complex geometries include more than one geographic item. 
(iii) The low definition of the terrain layer of the 3D maps.

\subsection{Discussion}
The results of the experiment suggest improving some aspects of the OTM user interface, but they positively answer our research questions.
Concerning RQ1, participants preferred the maps in which search results were clearly distinguishable from the background layer; e.g., OTM 3D ones. Moreover, they appreciated the opacity tuning as an information filtering function to support the reduction of visual complexity.
Concerning RQ2, results highlight the importance of transparency sliders to focus on useful data. We noticed that several participants used the sliders to hide irrelevant data categories. However, other people tuned opacity in a smoother way, maintaining semi-transparent data in the map.

\section{Conclusions}
\label{conclusions}

We presented a visualization model for representing Geographical Information search results on 2D/3D maps.
The model stylizes results, using their geometries, and represents them in vivid colors to enhance their discernibility. Moreover, it introduces transparency sliders to modify the opacity of data in the map.
A user test has shown that emphasizing search results through coloring, stylization and opacity tuning helps finding relevant information. Furthermore, transparency sliders are an efficacious tool to focus maps on specific information needs.
This work was funded by projects MIMOSA (``Progetto di Ateneo Torino\_call2014\_L2\_157'', 2015-17), "Ricerca Locale" and "Ricerca Autofinanziata" of the University of Torino.

\bibliographystyle{ACM-Reference-Format}

\end{document}